# Performance Analysis of a Mission-Critical Portable LTE System in Targeted RF Interference

Vuk Marojevic, Raghunandan M. Rao, Sean Ha, Jeffrey H. Reed
Bradley Department of Electrical and Computer Engineering
Wireless@Virginia Tech
Blacksburg, VA, USA
{maroje|raghumr|seanha65|reedjh}@vt.edu

*Abstract*—Mission-critical wireless networks are being upgraded to 4G long-term evolution (LTE). As opposed to capacity, these networks require very high reliability and security as well as easy deployment and operation in the field. Wireless communication systems have been vulnerable to jamming, spoofing and other radio frequency attacks since the early days of analog systems. Although wireless systems have evolved, important security and reliability concerns still exist. This paper presents our methodology and results for testing 4G LTE operating in harsh signaling environments. We use software-defined radio technology and open-source software to develop a fully configurable protocol-aware interference waveform. We define several test cases that target the entire LTE signal or part of it to evaluate the performance of a mission-critical production LTE system. Our experimental results show that synchronization signal interference in LTE causes significant throughput degradation at low interference power. By dynamically evaluating the performance measurement counters, the k-nearest neighbor classification method can detect the specific RF signaling attack to aid in effective mitigation.

## I. INTRODUCTION

Wireless infrastructure and technology add to the well-being of society by providing communications and multimedia services at an affordable cost. While the commercial sector continues to expand its service diversity, mission-critical networks and, in particular, public safety and military networks are looking to leverage advances in cellular technology and fully adopt the 4G long-term evolution (LTE) protocol.

Public safety units use wireless communication to effectively coordinate and provide assistance in time [1]. National security relies on sensors and wireless communications to efficiently assess and quickly respond to potential threats. The increasing number of unmanned vehicles poses more stress on reliable radio communications, where even a partial breakdown can have catastrophic consequences. Mission-critical systems, moreover, need to be quickly deployable and operated in non-ideal and even harsh signaling environments.

Wireless communication systems have been vulnerable to jamming, spoofing and other attacks since the early days of analog systems. Although wireless systems have evolved, important security and reliability concerns still exist. Different types of attacks to wireless networks have been the topic of research for several years [2] [3].

Lazos et al. [4] address the problem of control channel jamming in multi-channel ad-hoc networks and proposes a randomized distributed channel establishment scheme that allows nodes to select a new control channel using frequency hopping. Bicakci et al. [5] target practical hardware, software, and firmware solutions for 802.11 devices to efficiently combat Denial-of-Service (DoS) attacks. Chiang et al. [6] introduce a code-tree system for circumventing jamming signals. He et al. [7] show that controlled node mobility can be exploited for increasing the resilience against jamming.

References [8]–[11] investigate different types of RF attacks on LTE networks. Since LTE is an open standard, an adversary can generate a protocol-aware attack, where the interfering signal can be overlaid over a specific LTE physical channel to degrade the system performance at low profile. These papers conclude that relatively little energy is needed for causing major system performance degradation. Labib et al. [12] coin the term *LTE control channel spoofing*, which refers to transmitting the LTE synchronization signals or a partial LTE downlink (DL) control frame from a fake eNodeB (eNB). The experimental results show that this type of attack can cause DoS.

This paper analyzes the vulnerabilities of a mission-critical LTE system. We introduce a software-defined radio testbed and methodology for evaluating the impact of intentional RF interference on a production LTE network that is meant for mission-critical deployment. We provide experimental results to compare the effect of protocol-aware and unaware interference on LTE system performance. We conclude that interfering with LTE synchronization signals considerable degrades throughput with relatively low interference power. We also a method for protocol-aware interference detection based on k-nearest neighbor classification and evaluation of performance measurement (PM) counters. The rest of the paper is organized as follows. Section II presents the LTE system under test and briefly reviews the LTE control channels. Section III introduces our testbed and testing methodology. Section IV provides the performance results and analyses, whereas Section V illustrates the proposed detection mechanism. Section VI concludes the paper.

## II. MISSION-CRITICAL LTE SYSTEM UNDER TEST

The system that we analyze is a production LTE system built for military missions and next generation public safety trials. The system is embedded in a small form factor with the radio unit, the main unit and the power unit. The main unit features the Evolved Packet Core (EPC). This allows for rapid deployment in the field, needing only a power generator and an antenna mounted on a mast to establish a fully functional cell and offer LTE services. If backhaul is available, external networks can be accessed.

The above mission-critical LTE system adheres to the 3GPP LTE Release 8 specifications. That is, it creates LTE frames using the same control channels and signals as commercial LTE systems. Commercial UEs can be used with this network. This allows leveraging competitive R&D innovations and sophisticated handheld devices produced for the mass market and available at competitive prices. Note that the specifications for public safety LTE UEs differ from commercial UEs, allowing higher transmission power, among other features. The next generation public safety network, known as FirstNet in the US, requires compliance with 3GPP LTE Rel. 8 or higher. Our analysis does not assume any specific type of UE. We analyze the LTE network performance. Our results are generalizable across 3GPP compliant LTE networks and UEs since we do not assume any specific LTE-Advanced (Rel. 10 or higher) or LTE-Pro (Rel. 12 or higher) features.

The LTE control channels are essential for providing the capability for the rest of the system. Without control channels and signals, the rest of the communication network is unusable. Protocol-aware interference can target specific physical channels. We briefly review some of the fundamental LTE downlink (DL) and uplink (UL) control channels that are relevant for the experiments and analysis of this paper. These channels are available in all releases of LTE. Additional control channels or control information are needed for some of the more advanced LTE features, such as carrier aggregation and use of unlicensed spectrum.

*Primary and Secondary Synchronization Signals (PSS/SSS)*— The PSS/SSS need to be regularly tracked by the User Equipment (UE) in order to maintain synchronization with the eNB of the cell.

*Physical Broadcast Channel (PBCH)*—The PBCH contains the Master Information Block (MIB) which provides details about the downlink bandwidth, resource length of the Hybrid ARQ (HARQ) Indicator Channel (PHICH), and the System Frame Number (SFN) to aid the UE in frame synchronization. The PBCH is mapped to the central 72 subcarriers of the OFDM symbol and is spread over four frames. It is QPSK modulated with a 16-bit CRC, but with an aggregate coding rate of 1/48.

*Physical Downlink Control Channel (PDCCH)*—The PDCCH carries critical control information, such as UE resource allocation, Modulation and Coding Scheme (MCS) of user data, information about the HARQ and precoding matrices for MIMO. It is QPSK-modulated with rate 1/3 convolutional coding. During initial cell access, it informs the UE of the first System Information Block (SIB1). Without the SIB1, the UE will be unable to complete the cell attachment process. Additionally, after cell attachment, it would be impossible for the UE to decode its data if the PDCCH is improperly decoded.

*Physical Control Format Indicator Channel (PCFICH)*— The PCFICH contains information regarding the size of the PDCCH. It contains the Control Format Indicator (CFI), which is 2 bits, and is encoded using a block code rate of 1/16.

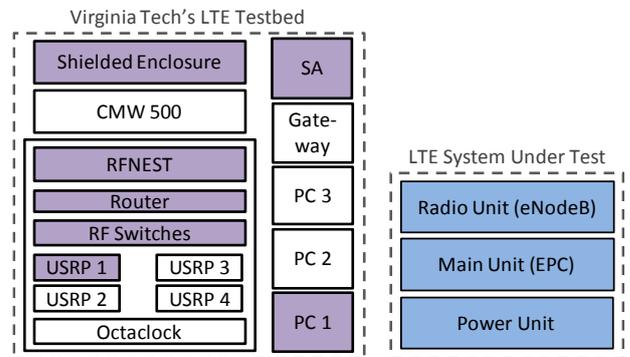

Fig. 1. Testbed hardware (shaded blocks are used in the experiments, SA: signal and spectrum analyzer).

*Cell-Specific Reference Signal (CRS)*—The CRS are downlink pilot symbols that are used to equalize the effects of the channel in order to perform coherent detection of the digitally modulated data symbols. CRS are QPSK-modulated and use a Gold sequence of length 31, which is initialized using the cell ID value. The CRS symbols are distributed sparsely in time and frequency, occupying about 5% of the REs.

*Physical Downlink Control Channel (PUCCH)*—Similar as the PUCCH, the PDCCH is a dedicated control channel that UEs use to request resources and provide related control information to the eNB.

*Physical Downlink and Uplink Shared Channels (PDSCH and PUSCH)*—These two channels carry the user data on the uplink and downlink along with certain control information, such as acknowledgements or negative acknowledgements. Note that when a user has an active data session, uplink control information is mapped to the PUSCH as opposed to PUCCH.

## III. LTE TESTBED AND TESTING METHODOLOGY

### A. LTE Testbed: Hardware

Virginia Tech has built an LTE testbed using software-defined radios, LTE test instruments and emulated (RFNEST) and real over-the-air channels [13]. The rackmount testbed has RF ports for attaching external RF signals. We use one of these ports for attaching the mission-critical LTE system and place the commercial UE in the shielded box. The UE operates receives and transmits over-the-air over a short distance inside the shielded box. The rest of the signal path goes through RF cables. Fig. 1 shows the test setup.

PC1 generates the interference waveform. The samples are passed to USRP1 via the Ethernet router. USRP1 creates the RF signal that goes into RFNEST via an RF switch. The purpose of the RF switch is to enable switching between channel emulation (RFNEST) and antenna (not shown here, see [13] for details). The interference RF signal is combined with the LTE signal from the eNodeB in RFNEST, which allows selecting independent gains/attenuations to obtain the desired interference to signal power ratio. The spectrum analyzer is used to empirically adjust the power levels as well as to ensure time synchronization needed for some of the test cases. Finally the combined signal is passed to an antenna inside the shielded enclosure, which also contains the UE. Note that the interferer also received the eNB downlink signal, in particular the PSS and SSS for synchronized interference strategies.

*B. LTE Testbed: Software*

The test methodology that we develop is based on testing the vulnerabilities of a system by analyzing the individual subsystems. By targeting a specific subsystem or a specific combination of subsystems at a time, we can evaluate the system performance and determine the weakest component in the system and revise it to improve the overall system robustness.

We propose a parametric framework for interference generation, using the same waveform as the target system. In the case of LTE, individual subcarriers and OFDM symbols can be toggled to rapidly generate wideband, narrowband, and protocol-aware interference over any section of the LTE signal. We used the open-source software library libLTE/srsLTE [16] and developed LTE protocol-aware interference waveforms that targets specific subcarriers and OFDM symbols. The library implements the LTE uplink and downlink waveforms and readily supports Commercial-off-the-shelf (COTS) SDR hardware.

*Asynchronous Interference Waveforms*—The asynchronous interference waveform generates interference on specific subcarriers. This type of interference can be of certain duration or continuous or discontinuous in time. We can use this setup to generate any interference to LTE that does not need time alignment with the LTE frame. In particular, we use it for generating full-band, partial-band, and Physical Uplink Control Channel (PUCCH) interference, but can also generate a bogus PSS and/or SSS signal (PSS/SSS spoofing) by replacing OFDM symbols with synchronization sequences. An example, a 1.4 MHz interference waveform with three blocks of active subcarriers is shown in Fig. 2.

*Synchronous Interference Waveforms*—Transmitting on top of specific physical channels on the downlink requires synchronization with the network to determine the location of the physical channels. Consequently, we use a setup where the interferer (1) acts as a receiver and synchronizes with the eNB, in this case, through LTE's PSS and SSS, and (2) synchronously transmits its interference payload. A configurable timing offset can be specified to account for transmission and other delays. Fig. 3 illustrates the synchronous interference waveform which targets the LTE PSS/SSS.

*C. Performance Evaluation Metrics*

In order to compare the vulnerabilities of different control channels, we define a uniform metric based on Interference to Signal power Ratio (ISR) values, control channel occupancy fraction in the LTE frame, and its relative power w.r.t. the data channels. In this regard, we define the following quantities: (a) Interference to Signal Ratio per Resource Element ($ISR_{RE}$), (b) Interference to Signal Ratio per Frame ($ISR_F$), and (c) Interference to Signal Ratio per Target Signal ($ISR_N$).

*Interference to Signal Ratio per Resource Element*—$ISR_{RE}$ is defined as the ratio of the interference signal power to that of the LTE signal, assuming that all the Resource Elements (REs) have the same transmit power.

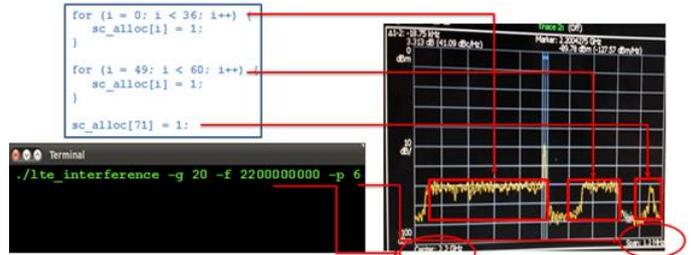

**Fig. 2.** Asynchronous interference waveform generation.

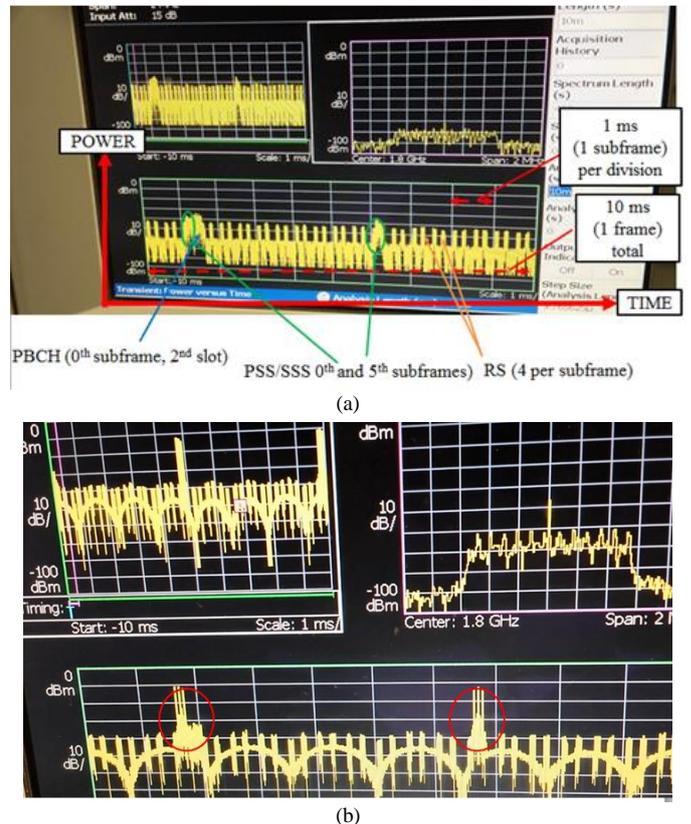

**Fig. 3.** Partial LTE DL signal which, for illustration purposes, consist of the PSS/SSS, PBCH, and CRS only (a). Partial LTE DL signal with synchronous PSS/SSS interference (b).

*Interference to Signal Ratio per Frame*—When the interferer targets a specific control channel, it occupies a specific fractions of the total number of REs in the LTE DL frame. To account for this, we define

$$ISR_F = \frac{ISR_{RE} \times N_{T,F}}{N_{tot,F}},$$

where $N_{T,F}$ denotes the number of REs targeted per frame, and $N_{tot,F}$ the total number of REs per frame. We use this metric to compare the effects of different interference strategies on system performance.

*D. Test Cases*

Table I presents the test cases or interference scenarios. The difference between PSS/SSS spoofing and interference is the following: In the case of spoofing, the attacker transmits a fake, but legitimate PSS/SSS non-synchronously to the LTE DL frame. PSS/SSS interference implies transmitting interference on top of the eNB's synchronization signals.

TABLE I. TEST CASES (INTERFERENCE SCENARIOS)

|   | Interference Scenario | Direction | Synchronous |
|---|---|---|---|
| 0 | No interference | - | - |
| 1 | Full-band interference | UL/DL | No |
| 2 | Half-band interference | UL/DL | No |
| 3 | PUCCH interference | UL | No |
| 4 | PUSCH interference | UL | No |
| 5 | PSS/SSS spoofing | DL | No |
| 6 | PSS/SSS interference | DL | Yes |

The interference node (PC1 with USRP1 in Fig. 1) uses the PSS/SSS from the eNB (LTE system under test) to synchronize the interference signal with the LTE frame at the UE. This is needed only for the test case 6. We used a 10 MHz FD-LTE signal in these experiments. The RF signal attenuators are electronically adjusted to achieve the desired ISR. For this we use RFview, the graphical user interface allowing digital control over all 8 signal paths of RFNEST [13]. The controlled test setup ensures a low-noise RF environment such that the LTE system performance becomes interference-limited.

## IV. EXPERIMENTAL RESULTS AND ANALYSES

We measure the UL and DL LTE throughput using iPerf to quantify the impact of interference. The results are shown in Figures 4 and 5.

The nominal LTE system throughput is around 12 and 8 mega-bits per second (Mbps) on the DL and UL, respectively. We observe that the throughput degrades as the interference covers more signal bandwidth. In other words, full-band interference is the most severe since all resource elements are affected. However, from Table II we see that this is not a power-efficient method since it requires a higher interference power.

PUSCH interference is the next most significant threat, but requires slightly less interference power, proportional to the span of the PUSCH w.r.t. the entire LTE system bandwidth. For 10 MHz FD-LTE PUSCH requires about 1.24 dB less power to cause the same degradation as full-band interference on the UL.

PSS/SSS spoofing does not have a significant effect on the throughput because, from the perspective of the receiver, the spoofing synchronization signals are simply asynchronous narrowband signals with a low duty cycle. However, synchronization signal spoofing impedes LTE network acquisition for UEs that are in the initial cell selection process, as demonstrated in [14] and [15]. For synchronous PSS/SSS interference, even with a high ISR, the interference does not cause synchronization loss; however, there is noticeable degradation of throughput, which proves to be a more serious and immediate threat than the potential loss of synchronization.

Because of the sparsity of resource elements that the PSS and SSS occupy in the LTE resource grid, synchronous PSS/SSS interference is a very energy-efficient interference strategy (Table II). PUCCH interference requires 20 times more energy to degrade the UL throughput just as much as PSS/SSS interference. However, the RF energy efficiency comes at the cost of higher complexity in the interference waveform generation because of tight synchronization requirements between the interferer and the UE.

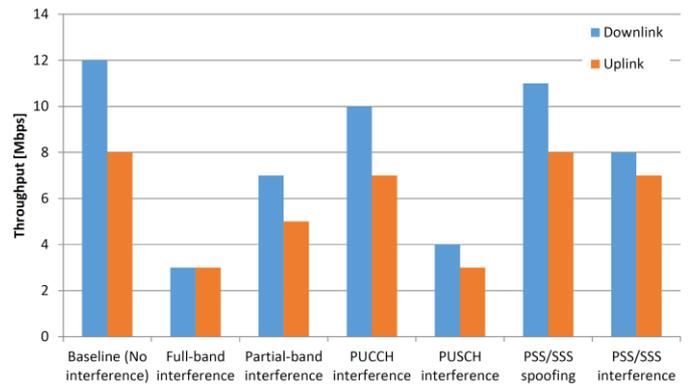

**Fig. 4.** Throughput results for $ISR_{RE} = 0$ dB.

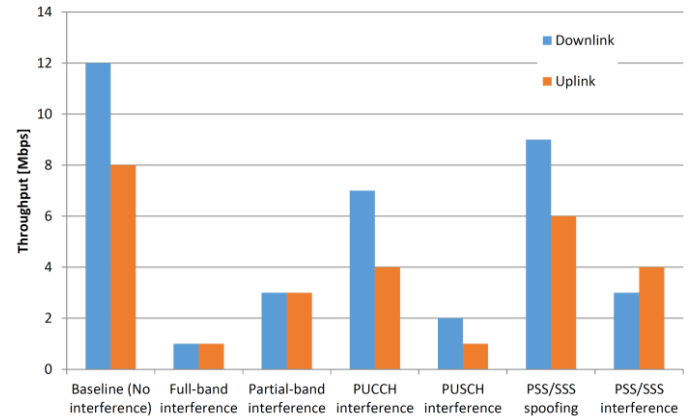

**Fig. 5.** Throughput results for $ISR_{RE} = 5$ dB.

TABLE II. RELATION BETWEEN $JSR_{RE}$ AND $JSR_F$

| Interference Scenario | $\left(\frac{N_{T,F}}{N_{tot,F}}\right)$ | $\frac{ISR_F}{ISR_{RE}}$ (dB) |
|---|---|---|
| Full-band interference | 100% | 0 |
| Half-band interference | 50% | -3.01 |
| PUCCH interference | 25% | -6.02 |
| PUSCH interference | 75% | -1.25 |
| PSS/SSS spoofing | 1.23% | -19.1 |
| PSS/SSS interference | 1.23% | -19.1 |

## V. INTERFERENCE DETECTION

The advantage of using mission-critical production LTE equipment is the ability to leverage sophisticated detection mechanisms to determine the presence of interference and determine the type of interference. The LTE test equipment that we used was equipped with *counters* which are typically used to measure system performance, but can also be leveraged to detect abnormal RF behavior. As an example, we present the case of PUCCH interference detection using a k-NN classification algorithm shown below.

Fig. 6 shows the 2-dimensional 3-Nearest Neighbor (3-NN) algorithm by monitoring two PUCCH-related performance metrics from our production LTE equipment, which we refer to here as *PM_Counter1* and *PM_Counter2*. For classifying a data point, we examine $k$=3 nearest data points surrounding it. The "blue cluster" in Fig. 6 denotes a classification of "Interference", whereas the "red cluster" denotes "No Interference". The

Algorithm 1: One iteration of k-NN classification

1. Initial inputs:
    $N$ metrics (PM Counters/ Key Performance Indicators) as feature-vector $[Metric_1, Metric_2, ..., Metric_N]$
    $n$ categories of classification $\{C_1, C_2, ... C_n\}$
    $M$ training samples as feature vectors: $\{m_1, ... m_N\}$, with each $m_i$ properly classified from one pf the $n$ possible categories.
2. Initialize training samples: $\{m_1, ... m_N\}$.
3. Input to current iteration of algorithm:
    Data point (as feature-vector) to classify $x = [Metric_1, Metric_2, ..., Metric_N]$
    For each $m_i$ in $\{m_1, ... m_N\}$
        Compute distance between $x$ and $m_i$: $d_i = distance(x, m_i)$.
4. Sort $\{d_1, ... d_N\}$ in order of increasing distance.
5. Select $\{m_1^*, ..., m_k^*\}$ as the $m_i$'s corresponding to the $k$ smallest entries of $\{d_1, ... d_N\}$.
6. Classify $x$ based on majority vote: $x$ belongs to the $C^*$ corresponding to the category that the majority of the $k$ training samples $\{m_1^*, ..., m_k^*\}$ belong to.

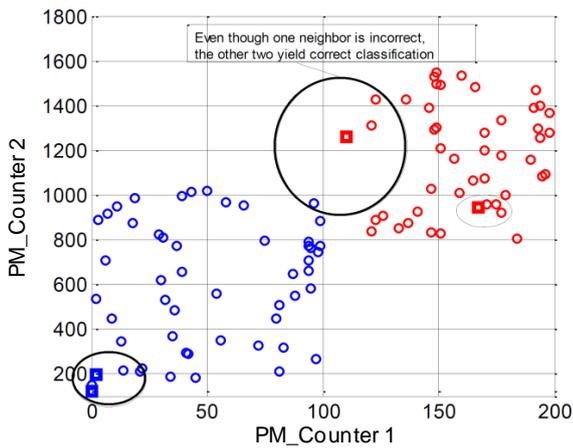

**Fig. 6.** Detection of PUCCH interference with a 3-NN classification algorithm, using two appropriate PM counters available in the production LTE eNB.

circles are dummy initialization points for the k-NN algorithm (may also represent training data) and the squares are actual data points gathered from our experiments. This example illustrates that k-NN is able to properly classify the given PM counter data, even though one data point deviates from the center of the pre-classified initialization points.

## VI. CONCLUSIONS

This paper has analyzed a mission-critical LTE system operating in a harsh signaling environment. The results have shown that PSS/SSS interference is a major threat to LTE performance after the UE attaches to a cell, and that full-band/half-band and PUSCH interference cause the most severe throughput degradation, but at the cost of higher power. We have also developed a k-NN clustering method that evaluates a subset of the available performance measurement counters to detect the nature of interference. These results demonstrate how existing mechanisms can be used to detect the presence of unusual interference in the network. This is a crucial first step for effective deployment and operation of 4G networks and designing interference-aware systems on the road to 5G. No wireless system can be made 100% secure and, at the same time, efficient. Hence, tradeoffs will need to be made when developing effective interference mitigation techniques. This is an important area in R&D that can significantly contribute to the evolution of wireless protocols towards 5G and beyond.